\documentclass[12pt]{article}
\usepackage{amssymb,amsmath,amsthm,graphicx,ulem}

\pdfoutput=1
\usepackage{hyperref}
\usepackage{graphicx,subfigure}
\usepackage{epsfig}
\usepackage{amsmath}
\usepackage{amsfonts}
\usepackage{amssymb}
\usepackage[usenames]{color}
\usepackage[letterpaper,left=2.cm,right=2.cm,top=2.5cm,bottom=2.5cm]{geometry}
\usepackage{fancyhdr}

\newcommand{\beq}{\begin{equation}}
\newcommand{\eeq}{\end{equation}}
\newcommand{\be}{\begin{equation}}
\newcommand{\ee}{\end{equation}}
\newcommand{\beqa}{\begin{eqnarray}}
\newcommand{\eeqa}{\end{eqnarray}}
\newcommand{\beqar}{\begin{eqnarray*}}
\newcommand{\eeqar}{\end{eqnarray*}}
\newcommand{\bea}{\begin{eqnarray}}
\newcommand{\eea}{\end{eqnarray}}

\newcommand{\ra}{\rightarrow}
\newcommand{\om}{\omega}
\newcommand{\si}{\sigma}

\newcommand{\p}{\partial}

\newcommand{\ka}{\kappa}

\newcommand{\lp}{\left(}
\newcommand{\rp}{\right)}

\newcommand{\dd}{\textrm{d}}

\numberwithin{equation}{section}

\newcommand{\nn}\nonumber
\newcommand{\eqn}[1]{(\ref{#1})}

\numberwithin{equation}{section}

\begin{document}

\allowdisplaybreaks

\normalem

\title{A Simple Holographic Model of Nonlinear Conductivity}

\author{Gary T. Horowitz${}^{\,a}$, Nabil Iqbal${}^{\,b}$, Jorge E. Santos${}^{\,a,\,c}$
\\
\\
${}^{\,a}$ Department of Physics, UCSB, Santa Barbara, CA 93106, USA
\\
${}^{\,b}$ Kavli Institute for Theoretical Physics, UCSB, Santa Barbara CA 93106
\\
${}^{\,c}$ DAMTP, Centre for Mathematical Sciences, University of Cambridge,
\\
 Wilberforce Road, Cambridge CB3 0WA, UK
\\ 
\\
 \small{gary@physics.ucsb.edu, niqbal@kitp.ucsb.edu, jss55@physics.ucsb.edu}
 }

 \date{}

\maketitle 

\thispagestyle{fancy}
\fancyhead{}
\fancyfoot{}
\renewcommand{\headrulewidth}{0pt}
\rhead{NSF-KITP-13-215}

\begin{abstract}
\noindent  We present a simple analytic gravitational solution which describes the holographic dual of a $2+1$-dimensional conductor which goes beyond the usual linear response. In particular it includes Joule heating. We find that the nonlinear frequency-dependent conductivity is a constant. Surprisingly, the pressure remains isotropic. We also apply an electric field to a holographic insulator and show that there is a maximum electric field below which it can remain an insulator. Above this critical value, we argue that it becomes a conductor due to pair creation of charged particles. Finally, we study $1+1$ and $3+1$ dimensional conductors at the nonlinear level; here exact solutions are not available and a perturbative analysis shows that the current becomes time dependent, but in a way that is captured by a time-dependent effective temperature. 
 \end{abstract}

\newpage


\tableofcontents
\baselineskip16pt

\section{Introduction}

Motivated by gauge/gravity duality, there has been growing interest in dual (or holographic) gravitational descriptions of condensed matter phenomena. One of the most basic quantities of interest in condensed matter is the conductivity. This is usually computed holographically using linear response. One starts by finding a suitable stationary black hole solution to Einstein-Maxwell theory (with negative cosmological constant and possibly additional matter fields) to describe the equilibrium configuration. Then one perturbs this solution with boundary conditions at infinity corresponding to adding a homogeneous electric field to the dual system. Imposing ingoing waves at the horizon, one solves for the perturbation and reads off the current from the subleading term in the perturbation. The conductivity is then just the ratio of the current to the applied electric field. In most cases, both the background black hole and the linearized perturbation must be found numerically.

In this note, we present a simple model in which the black hole and perturbation can be solved analytically. Furthermore, the backreaction can be taken into account analytically so that one obtains the nonlinear effects of the conductivity. In particular, we see the temperature of the  black hole increase due to Joule heating.\footnote{Strictly speaking, the temperature of a time dependent black hole is not well defined. We will see that the mass of the black hole increases at just the rate expected from the increase in internal energy due to Joule heating.} We also present exact solutions which describe the application of an electric field to the gravitational dual of an insulator. These solutions only exist up to a maximum electric field. Above this value, we argue the solution collapses down to our time dependent black hole.  In the dual theory, this has a simple interpretation: At sufficiently high electric fields, one expects pairs of charged particles to be created and a current to be generated.

For a constant electric field, we find that the current remains constant and does not increase with time despite the fact that the temperature changes. More curiously, even though we apply an electric field in one direction and generate a current in that direction, the pressure remains isotropic. Our examples are ($3+1$)-dimensional gravitational solutions which are dual to ($2+1$)-dimensional systems, and we will show that these features can be understood in terms of electromagnetic duality in the bulk. In other dimensions, there are no simple exact solutions, but we will present  perturbative calculations which show that the current does change with time and the solution is no longer isotropic when $D \ne 4$.

\section{Nonlinear holographic models in $D=4$}
\subsection{A conductor}

We consider four-dimensional 
 Einstein-Maxwell theory with negative cosmological constant:
\be\label{action}
S = {1\over 16\pi G}\int d^4 x \sqrt{-g} (R + 6 - F^2)
\ee
where we have set the AdS radius to one.
Consider the four dimensional planar Schwarzschild-AdS black hole: 
\be
ds^2 = -f(r) dt^2 + {dr^2\over f(r)} + r^2 (dx^2 + dy^2)
\label{eq:lineelement4D}
\ee
with $f = r^2 - r_0^3/r$. This is dual  to a $2+1$ dimensional CFT at temperature $T = 3r_0/4\pi$. The conductivity, computed using linear response as
\begin{equation}
\sigma(\omega)\equiv\frac{J(\omega)}{E(\omega)}=\frac{1}{4\pi G}\lim_{r\to+\infty}r^2\frac{F_{xr}}{F_{xt}}\,
\label{conductivity}
\end{equation}
is well known to be constant\footnote{The factor of $G^{-1}$ is proportional to the number of degrees of freedom in the CFT, and we keep it explicit to simplify the discussion of the Joule heating. If we write the coefficient of $F^2$ as $1/(4\,g^2_F)$, Eq.~(\ref{eq:conductivityresult}) corresponds to $\sigma = 1/g^2_F$, as presented in \cite{Herzog:2007ij}.} \cite{Herzog:2007ij}
\begin{equation}
\sigma(\omega) = \frac{1}{4\pi G}\,.
\label{eq:conductivityresult}
\end{equation}

Normally, a translationally invariant system has infinite DC conductivity since momentum conservation means that the charge carriers cannot dissipate their momentum. One exception to this is if there is zero net charge density, as is the case here.  Under an applied electric field, the positive and negative charges move in opposite directions. They both contribute positively to the current, but cancel out in the momentum. So the current and momentum essentially decouple.

Rather than taking the limit $\omega \rightarrow 0$ of $\sigma(\omega) $, we start by  computing the DC conductivity directly, at linear order. We want to apply a uniform electric field on the boundary so we set $F_{xt} = E$. It is easy to check that this satisfies Maxwell's equations, but it is singular on the horizon: $F_{\mu\nu} F^{\mu\nu} = -E^2/fr^2$ . To find  a solution that is regular on the future horizon (corresponding to ingoing wave boundary conditions) we introduce ingoing Eddington coordinates.
Setting $dr_* = dr/f$, and $v = t + r_*$, the metric (\ref{eq:lineelement4D}) becomes
\be
ds^2 = -f(r) dv^2 + 2dv dr + r^2 (dx^2 + dy^2)
\ee
A solution to Maxwell's equations that is regular in these coordinates and includes $F_{xt} = E$ is simply 
\be\label{maxwell} 
F_{xv} = E
\ee
 Re-expressing this solution in terms of the original coordinates, we find that in addition to $F_{xt} = E$, we also have $F_{xr} = E/f$. This means that we have induced a current in the dual theory.  From (\ref{conductivity}) we see immediately that $\si_{DC} = 1/(4 \pi G)$.

We now want to go beyond this linear calculation and include the backreaction of the Maxwell field on the metric. This is easily done in ingoing Eddington coordinates.  The key point is that the Maxwell field \eqn{maxwell} is null\footnote{Note that if $k$ is the null vector $\p/\p r$, then $k_\mu = \p_\mu v$.}: $F_{\mu\nu} F^{\mu\nu} = 0$. The only nonzero component of the stress tensor is $T_{vv} = E^2/r^2$, which corresponds to an ingoing flux of null energy. The exact solution of Einstein's equation with a null stress tensor of this form is well known. It is the Vaidya metric:
\be\label{vaidya}
ds^2 = -\left[ r^2 - {2\,m(v)\over r}\right] dv^2 + 2dv dr + r^2 (dx^2 + dy^2)
\ee
where $m(v) = (r_0^3/2) + E^2 v$.
This corresponds to a black hole whose mass grows linearly in time. The Maxwell field in the exact solution is  the same as the linearized solution, \eqn{maxwell}. From the standpoint of the dual field theory, this means that the exact DC conductivity is still $\si_{DC} = 1/(4\pi G)$. Since $m$ is a measure of the mass density (for a static black hole: $M/V = m/(4\pi G)$), the black hole mass increases at exactly the rate expected from Joule heating: $dM/dt = \vec E \cdot \vec J$, so $dm/dt = E^2$. This is a consequence of the fact that the holographic stress tensor satisfies
\be
\p_\mu T^{\mu\nu} = F^{\mu\nu}J_\mu
\label{eq:ward}
\ee

This solution is easily generalized to allow time dependent electric fields on the boundary. One simply takes $F_{xv}= E(v)$ and  the metric given by \eqn{vaidya} with $m(v) = r_0^3/2 + \int E(v)^2 dv$. This means that we can holographically describe the process of turning on an electric field, keeping it constant for a while, and then turning it off again. It also means that we can apply an oscillating electric field of any frequency.  Since $F_{xv}= E(v)$  implies that $J(v) = E(v)/(4\pi G)$ we see that $\sigma(\omega) = 1/(4\pi G)$, even nonlinearly. One can understand the simplicity of the conductivity using electromagnetic duality in the bulk \cite{Herzog:2007ij,Hartnoll:2007ip}. This will be discussed further in section 4.

This solution has a current, but no momentum. As we mentioned above, this is expected since we have started with no net charge density. 
A more puzzling feature of our solution is the fact that the metric remains isotropic in the $x,y$ directions,  so the pressures are equal: $T_{xx} = T_{yy}$. Even though we apply an electric field in the $x$ direction and generate a current in the $x$ direction, the pressures remain equal!

We will see in the next section that  this is a special feature of four dimensional bulk spacetimes, or $2+1$ dimensional boundary theories.  It can be viewed as another consequence of electromagnetic duality. Under $F\ra *F$, the metric does not change, but the Maxwell field rotates to $F_{yv} = E$ which corresponds to having an electric field and current in the $y$ direction. This means that the pressures must be isotropic. We will discuss a possible microscopic explanation for this in section 4.

 \subsection{An insulator}

We now ask what happens if one applies a constant electric field to a confining vacuum. A simple example of such a vacuum is given by the AdS soliton \cite{Witten:1998zw,Horowitz:1998ha}, which is just the double analytic continuation of the planar black hole. To obtain a fully backreacted solution with an electric field, we simply start with the magnetically charged planar black hole. This has a Maxwell field $F_{xy} = B$, so if we analytically continue $y = i\tau$, $t = i \theta$, and $B = - iE$,  we get 
\be\label{soliton}
ds^2 = f(r) d\theta^2 + {dr^2\over f(r)} + r^2 (dx^2 -d\tau^2)
\ee
with 
\be f(r) = r^2 - {2m_0\over r} - {E^2 \over r^2}
\ee
The radial coordinate only exists for $r \ge r_+$ where  $r_+$ is the root of $f$. Regularity at the tip, $r=r_+$, requires that $\theta$ be periodic with period $L = 4\pi/f'(r_+)$.  The Maxwell field is simply $F_{x\tau} = E$ which is nonsingular everywhere. This solution corresponds to no current in the dual theory, so as expected it describes an insulator. 

However, for fixed $L$ this solution only exists up to a maximum value of the electric field. Expressing $E$ in terms of $r_+$ and $L$ we obtain
\be\label{Efield} 
E^2 = \lp {4\pi\over L} - 3r_+\rp r_+^3
\ee
The original AdS soliton with $E=0$ has $r_+ = 4\pi/3L$. As we increase the electric field, $r_+$ decreases.
The maximum value of $E$ occurs at $r_+ = \pi/L$ where $E=E_{crit} \equiv (\pi/L)^2$. For each $E<E_{crit}$ there are two possible values for $r_+$. One might have thought that only the larger value yields a solution since once the circle pinches off,  the spacetime ends. However the two values of $r_+$ do not correspond to two roots of the same function $f(r)$, but rather the roots of two $f$'s with different $m_0$. Thus there are two solitons for each $E<E_{crit}$.\footnote{This is analogous to the fact that there are two spherical AdS black holes with the same temperature, provided $T$ is larger than a critical value.}

These two solutions have different mass, and the one with largest $r_+$ has the lowest mass. If we compactify $x$ with period $\tilde L$ to make the total mass finite, we have 
\be
M_{sol} = - {m_0  \over 8\pi G} L \tilde L \ . \label{eq:solitonmass}
\ee
Note the overall minus sign. When $m_0 >0$, the soliton has lower energy than pure AdS.  The mass parameter $m_0$ can  be expressed in terms of $r_+$ and $L$: 
\be
m_0 = 2r_+^3 - {2\pi r_+^2\over L}
\ee
At the critical electric field $r_+ = \pi/L$, and $m_0 = 0$. So adding an electric field to the AdS soliton increases its mass and at the maximum value of $E$, the mass vanishes.

What happens if one increases the electric field past this critical value? One possibility is that there is another branch of static solutions which break some of the translational symmetry and support larger values of $E$. We have found no evidence for this. Furthermore, a static soliton  which is nonuniform in the $x$ direction could be analytically continued  to produce a nonuniform planar black hole. Solutions of this type are known to exist if one includes an axion field (see, e.g. \cite{Donos:2011bh}), but are widely believed not to exist in pure four dimensional  Einstein-Maxwell theory.

It is much more likely that there is no static solution with $E>E_{crit}$. In this case, if one increases $E$ greater than $E_{crit}$ the soliton will presumably collapse down to form the above black hole. This solution has a current and the interpretation is simply that the electric field is now strong enough to break apart neutral confined massive states, releasing charged and deconfined matter that can carry the current. Recall that in the dual theory, the mass of all excitations (and thus the binding energy to overcome) is $O(1/L)$.

 For $ E < E_{crit}$, in the field theory one might expect a small current caused by the pair creation of massive charged glueballs. In the gravity description this is dual to the Schwinger pair creation of the quanta of charged {\it bulk} fields. This leads to a current that is $O(1)$ in the large $N$ limit but is exponentially suppressed in the electric field.  

It is also interesting to consider the situation at a finite temperature $T$. Consider first starting with the soliton with zero electric field. If we increase the temperature above $T = 1/L$,
there is a  well-known phase transition in which the thermodynamically preferred solution changes from a gas of particles in the soliton background to a black hole; this is the Hawking-Page transition \cite{Witten:1998zw,Hawking:1982dh}. There is a sense in which this phase transition is connected  (along a line of first order phase transitions) to  the transition at $E = E_{crit}$  (and $T=0$) described above. The idea is that if we apply an electric field to the soliton, its mass becomes less negative, and so does its free energy,  $F = M_{sol}$. The free energy of the planar black hole is also negative, and becomes less negative as we decrease $T$. So for each $E < E_{crit}$ one should be able to find a $T$ so that the free energies are equal. This critical line $T(E)$ includes  $E = E_{crit}$, $T=0$ since both free energies vanish there. However, it is hard to make this precise because the black hole under consideration is time dependent, and thus not in thermal equilibrium.

\section{Generalizations to $D\ne 4$}

 Our simple holographic model does not extend to $D\ne 4$.
 While the DC conductivity at linear order can be easily calculated in any dimension, there does not appear to be an  exact solution for the full nonlinear conductivity. In this section, we will proceed perturbatively, and compute the current to third order in the electric field. While the results remain analytic, they cannot reasonably be called ``simple".

We consider $D$-dimensional Einstein-Maxwell theory with negative cosmological constant:
\begin{equation}
S = {1\over 16\pi G}\int d^D x \sqrt{-g} \left[R +{(D-1)(D-2)} - F^2\right]\,,
\label{eq:einsteinmaxwellD}
\end{equation}
where we have again set the AdS radius to one.
Our $D$-dimensional ansatz takes the following form:
\begin{subequations}
\begin{align}
&\dd s^2 = -f(v,r)\dd v^2+2 dv\,\dd r+\Phi_1(v,r)\, r^2\,\dd x^2+\Phi_2(v,r)\,r^2\dd \mathbf{y}^2_{D-3} \label{eq:ansatzmetric}\\
&F = Q_1(v,r) \dd x \wedge \dd v+Q_2(v,r) \dd x \wedge \dd r\,,
\label{eq:ansatzgauge}
\end{align}
\end{subequations}
where the last term in Eq.~(\ref{eq:ansatzmetric}) represents a $(D-3)$-dimensional flat space and is absent in $D=3$. The above line element still has residual gauge freedom, namely
\begin{equation}
{r}\rightarrow {r}+\gamma(v)\,.
\label{eq:gaugefreedom}
\end{equation}
We fix this gauge freedom by demanding that, for all $v$, the apparent horizon is located at a fixed radial distance $r=r_+$.

We then expand the gauge field and metric functions as:
\begin{equation}
f(v,r)=\sum_{j=0} \mathcal{E}^{2j}\,f^{(j)}(v,r)\,,\quad \Phi_i(v,r)=\sum_{j=0} \mathcal{E}^{2j}\,\Phi_i^{(j)}(v,r)\quad\text{and}\quad Q_i(v,r)=\sum_{j=0} \mathcal{E}^{2j+1}\,Q_i^{(j)}(v,r)\,,
\label{eq:perturbations}
\end{equation}
where $i=1,\,2$, and $\mathcal{E}$ is our expansion parameter. The metric functions are even in $\mathcal{E}$ because the Maxwell stress energy tensor is quadratic in the field strength, and the background solution we start with is neutral. We will later see that $\mathcal{E}$ is closely related to the field theory electric field $E$.

For the background ($j=0$) we take the $D$-dimensional AdS-Schwarzschild black hole with planar horizon topology:
\begin{equation}
f^{(0)}(r)=r^2-\frac{r_+^{D-1}}{r^{D-3}}\quad\text{and}\quad\Phi^{(0)}_{i}(r)=1\,.
\label{eq:schwarzchild}
\end{equation}

\begin{figure}[ht]
\centering
\includegraphics[width=.4\textwidth]{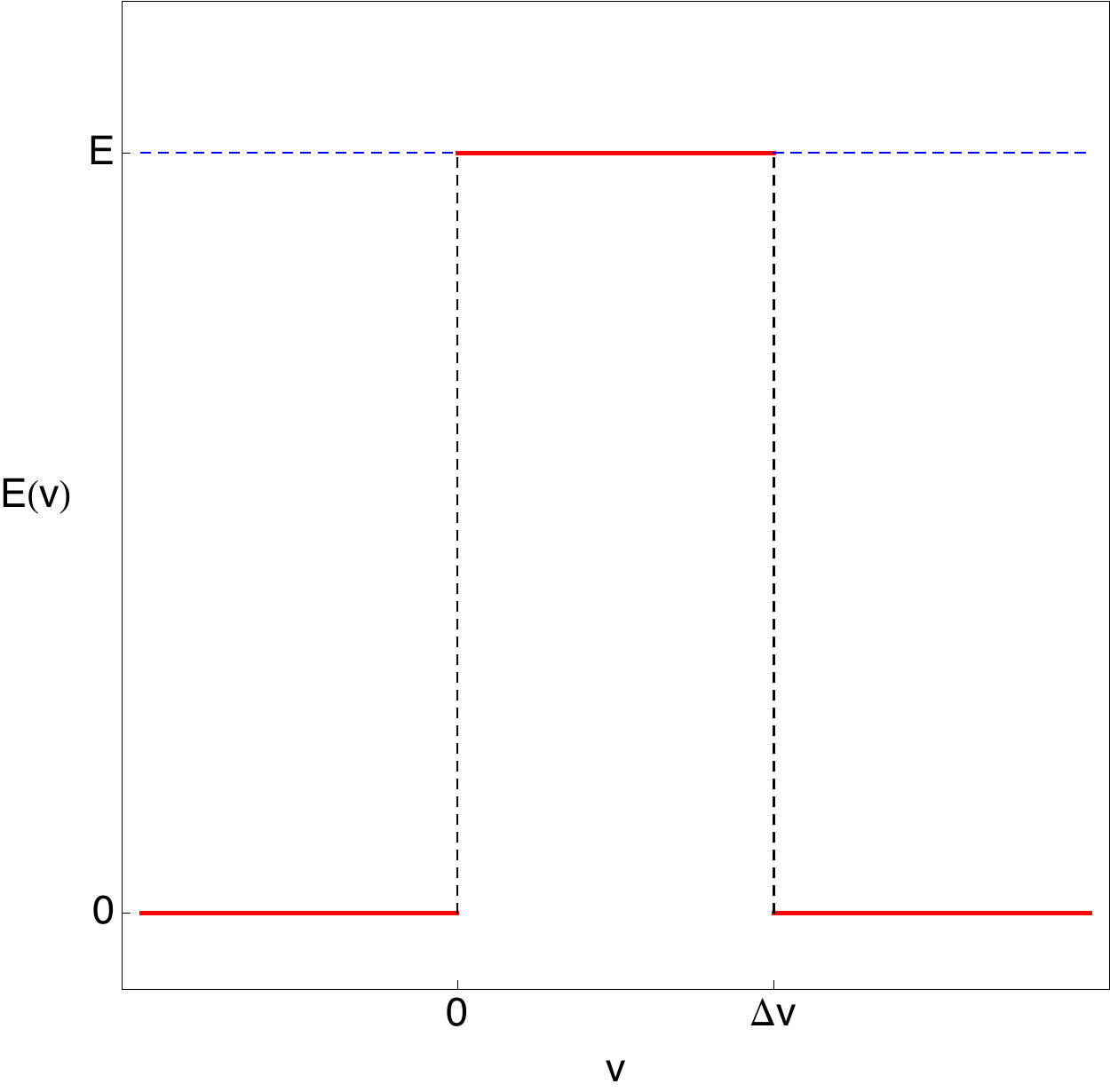}
\caption{Sketch of the electric field profile as a function of $v$.}
\label{fig:electricfield}
\end{figure}

We now set up the boundary electric field as shown in Fig.~\ref{fig:electricfield}. The electric field is assumed to be zero for $v<0$, in which case the metric is described by Eq.~(\ref{eq:schwarzchild}) and the Maxwell field strength is zero. At $v =0$ the perturbative field strength turns on, and it stays constant until we reach $v=\Delta v$, when it is turned off. We shall see later that our perturbation scheme further requires $\Delta\,v$ to be bounded above by $E^{-2}$.

At linear order, $j=1$, we can readily integrate Maxwell's equations for arbritrary $D$:
\begin{equation}
Q^{(1)}_1 = C_0,\quad Q^{(1)}_2 = \frac{1}{f^{(0)}( r )}\left[\frac{C_1}{r^{D-4}}-C_0\right]\,,
\label{eq:linearmaxwell}
\end{equation}
where both $C_0$ and $C_1$ are constants of integration. We fix $C_1$ in terms of $C_0$ by demanding regularity at the horizon: $C_1 = C_0\,r^{D-4}_+$. $C_0$, in turn, is fixed by demanding the field theory electric field along the $x-$direction to be $E$. For the case in which both the electric field and the current are constants, one can compute $C_0$, thus determining the relation between the applied electric field and $\mathcal{E}$, and also read the concomitant current. For all $D$, we find
\begin{equation}
C_0=1,\quad E\equiv\lim_{r\to+\infty}F_{xt} = \mathcal{E},\quad \text{and}\quad 4\pi G\,J\equiv \lim_{r\to+\infty}r^{D-2}F_{xr}\,=\, \mathcal{E}\,r^{D-4}_+,
\label{eq:currentlinear}
\end{equation}
where we have used that, for the time independent background (\ref{eq:schwarzchild}), $\dd v = \dd t +\dd r/f^{(0)}(r)$. This also allow us to write the conductivity, defined as the ratio of the current to the applied electric field, solely in terms of the background temperature $T^{(0)}_H\equiv (D-1)r_+/(4\pi)$:
\begin{equation}
\sigma^{(0)}_{DC}\equiv \frac{J}{E}=\frac{1}{4\pi G}\left[\frac{4\pi T^{(0)}_H}{D-1}\right]^{D-4}\,.
\label{eq:conductivitylinear}
\end{equation}
This is expected since the conductivity  has dimensions when $D \neq 4$, and the only scale in the background is the temperature. For $D=4$ we recover the familiar result: $\si_{DC}=1/(4\pi G)$. 

We now turn our attention to the nonlinear corrections to this expression. Because the equations become increasingly cumbersome to solve as one goes up in dimension, we will consider specific dimensions from now on, namely $D=3$ and $D=5$.

\subsection{Nonlinear $DC$ conductivity in $D = 3$}
Einstein-Maxwell theory in three space-time bulk dimensions is special, as bulk gauge fields tend to have logarithmic divergences. From the point of view of the dual field theory, this arises from the fact that a current in $1+1$ dimensions has dimension $1$, and thus the boundary double-trace coupling $\frac{1}{\ka} J^{\mu}J_{\mu}$ is marginal and runs logarithmically. This must be taken into account via modified boundary conditions, which were discussed in detail in \cite{Faulkner:2012gt}. 

To do this one introduces a radial cut off scale $r_{\Lambda}$, which is used to define the boundary electric field as:
\begin{equation}
F_{\mu\nu}(r_{\Lambda}) +\frac{8\pi G}{\kappa}\partial_{[\mu} J_{\nu]}= F^{\mathrm{source}}_{\mu\nu}\,,
\label{eq:boundary3D}
\end{equation}
where Greek indices run through boundary directions only, \emph{i.e.} $\mu=t,\,x$. $F(r_{\Lambda})$ is the bulk electric field evaluated at $r_{\Lambda}$, and $F^{\mathrm{source}}$ is the CFT electric field source. This boundary condition is of the usual form for a CFT double-trace deformation; it mixes the  source in the undeformed theory $F(r_{\Lambda})$ with the response $j_{\mu}$ to give the true field theory source $F^{\mathrm{source}}$. 

Once the boundary condition Eq.~\eqref{eq:boundary3D} is imposed, there are no longer any logarithmic divergences. Instead, due to the presence of the marginal coupling $\ka$, dimensional transmutation occurs and all physical quantities can be expressed in terms of the RG-invariant combination $r_{\star} = r_{\Lambda}\exp\left(-\frac{1}{\kappa}\right)$. $r_{\star}$ is precisely analogous to $\Lambda_{QCD}$ in the context of four-dimensional gauge theory. From an operational point of view, the fact that all physics can be expressed in terms of $r_{\star}$ is a rather nontrivial check of our calculations. 

Because we are interested in electric fields applied in the $x-$direction, the only nonzero component in $F^{\mathrm{source}}_{\mu\nu}$ is $F^{\mathrm{source}}_{x\,t}=E$. Note that our linear result for arbitrary dimensions, Eq.~(\ref{eq:conductivitylinear}) together with Eq.~(\ref{eq:boundary3D}) indicates that at the linear level $F^{\mathrm{source}}_{\mu\nu} = F_{\mu\nu}(r_{\Lambda})$ when $D=3$.

Our first task is to solve for $\Phi^{(2)}_1(v,r)$ and $f^{(2)}(v,r)$, for which one finds
\begin{subequations}
\begin{align}
& f^{(2)}(v,r)=\frac{2}{r_+^3}\left[2(r-r_+)(1+r_+ v)-r_+\log\left(\frac{r}{r_+}\right)\right]\,
\\
& \Phi^{(2)}(v,r)=\frac{4}{r^4_+}\left[\frac{v\,r^2_+}{r}-\log\left(\frac{r}{r+r_+}\right)\right]\,.
\label{eq:secondorder3D}
\end{align}
\end{subequations}
At first sight, these equations do not seem to correspond to asymptotically AdS solutions. However, this is an illusion caused by the gauge we choose, in which $f(v,r_+)=0$ to all orders in $\mathcal{E}$. In order to prove that this is the case, we have explicitly mapped the $(v,r)$ coordinates into Fefferman-Graham coordinates to $\mathcal{O}(z^2)$. The explicit map is given by
\begin{subequations}
\begin{align}
&r =\frac{1}{z}+\frac{r_+^2 z}{4}+\mathcal{E}^2\left[-\frac{2 \left(r_+\,t +1\right)}{r_+^3}+\frac{z \left(4 r_+ t -2 \log r_++9\right)}{4 r_+^2}-\frac{z \log (z)}{2 r_+^2}\right]+\mathcal{O}(z^2,\mathcal{E}^4)
\\
&v =t-z -\frac{r_+^2 z^3}{12}+\mathcal{E}^2\left[\frac{z^3 \log z}{6 r_+^2}-\frac{z^3 \left(12 r_+ t-6 \log r_+-1\right)}{36 r_+^2}\right]+\mathcal{O}(z^4,\mathcal{E}^4)
\end{align}
\end{subequations}
This expansion also allows us to extract the 2-dimensional holographic stress energy tensor up to second order in $E$:
\begin{equation}
4\pi G\,T_{ab}=\left[
\begin{array}{cc}
\frac{r_+^2}{4} + \frac{E^2}{2r_+^2} \left[2(r_+\,t-1)+\log\left(\frac{r_\star}{r_+}\right)\right]& 0
\\
\\
0 & \frac{r_+^2}{4} + \frac{E^2}{2r_+^2} \left[2r_+\,t-1+\log\left(\frac{r_\star}{r_+}\right)\right]
\end{array}
\right]+\mathcal{O}(E^4)\,.
\label{eq:stresstensor3D}
\end{equation}
In order to derive the holographic stress energy tensor we had to add to the on shell action the following counter terms
\begin{equation}
S_c = -\frac{1}{4\pi G}\int_{\partial\mathcal{M}} \dd x^2\sqrt{-\gamma}\,-\frac{8\pi G}{\kappa}\int_{\partial\mathcal{M}} \dd x^2\sqrt{-\gamma}\,J_\mu J^{\mu},
\label{eq:counter3D}
\end{equation}
where $\partial \mathcal{M}$ is the boundary metric and $\gamma$ its induced metric. The former term has been extensively used in the literature after the seminal work of \cite{Balasubramanian:1999re}, whereas the latter is associated with the interpretation given in \cite{Faulkner:2012gt}.

From Eq.~(\ref{eq:stresstensor3D}) one can see that perturbation theory will break down if $t$ is too large. This is the reason why $\Delta\,v$ in Fig.~\ref{fig:electricfield} should be bounded above by $E^{-2}$, such that the $\mathcal{O}(E^2)$ term in Eq.~(\ref{eq:stresstensor3D}) is still small when compared to its background value for all $v$.

We now go to third order, and determine $Q^{(3)}_i$ using the previous expansions. The result is:
\begin{subequations}
\begin{equation}
Q^{(3)}_1(v,r) = -\frac{2}{\kappa\, r_+^4 (r+r_+)} \left[\kappa  \left(r+r_+\right) \log \left(\frac{r+r_+}{r_{\Lambda }}\right)-2 \kappa\,r_++r+r_+\right]
\end{equation}
\begin{multline}
Q^{(3)}_2(v,r) = \frac{2}{r_+^5 (r_+-r) (r+r_+)^2} \Bigg\{\, r^2 \log \left[\frac{r_+}{r}\frac{r+r_+}{r_\Lambda}\right]+\,r_+^2 \log \left[\frac{r_+}{r} \frac{r_\Lambda}{r+r_+}\right]+\\
\,r_+\,r \log \left(\frac{r_+}{r}\right)+(r-r_+)\left[v\,(r+3r_+)r_++(r+r_+)\frac{1-\kappa}{\kappa}\right]\Bigg\}\,.
\end{multline}
\end{subequations}
This allow us to arrive at an expression for the boundary current to third order in the boundary electric field $E$:
\begin{equation}
4\pi G\,J = \frac{E}{r_+}\left\{1-\frac{2\,E^2}{r_+^4}\left[r_+ t-1-\log\left(\frac{\,r_{\star}}{r_+}\right)\right]\right\}\,.
\label{eq:current3D}
\end{equation}
One can easily check that with $J$ defined in Eq.~(\ref{eq:current3D}), and with $T_{\mu\nu}$ defined in Eq.~(\ref{eq:stresstensor3D}), the Ward identity Eq.~(\ref{eq:ward}) is readily obeyed.
Note that the current is now explicitly time dependent. 

In a time dependent setup, it is hard to define a temperature. Nevertheless, for small $E$, the black hole evolves slowly and it is natural to define an effective temperature via
\begin{equation}
T_H(v)\equiv \frac{1}{4\pi}\left|\frac{\dd f(v,r)}{\dd r}\right|_{r = r_+}\, .
\label{eq:effectivetemperature}
\end{equation}
 We  claim that it is natural to adopt this definition even when $E$ is not small and the black hole is not varying slowly. This is because  it is clear from our exact solution in $D=4$ that if we change $E$ abruptly at the boundary, the black hole responds at the same value of $v$. In other words, the influence of the source propagates in along an ingoing null surface.

Using the definition of effective temperature (\ref{eq:effectivetemperature}), we can express the time dependence of the stress tensor fully in terms of an effective temperature:
\be
4\pi G\, T_{\mu\nu} = \pi^2 T_H(t)^2 \Delta_{\mu\nu} + E^2 \tau_{\mu\nu} + \mathcal{O}(E^4)
\ee 
Here $\Delta_{\mu\nu} \equiv \mbox{diag}(1,1)$. The first term is the usual stress tensor for a 2d CFT in thermal equilibrium at a temperature $T$. The second term is a time-independent $O(E^2)$ correction whose precise form may be read off from \eqref{eq:stresstensor3D}. 

One can also express  the $DC$ conductivity ($J/E$) to second order in $E$ in terms of this effective temperature:
\begin{equation}
4\pi G\,\sigma_{DC} = \frac{1}{2\pi T_H(t)}\left[1+\frac{2E^2}{r_+^4}\log\left(\frac{e^{3/2}r_{\star}}{r_+}\right)\right]+\mathcal{O}(E^4)\,. \label{eq:conductivity3D}
\end{equation}
We can  see from the above equation that we cannot take the zero temperature limit, since this would imply $r_+\to 0$, in which case our last term in the above expression becomes comparable with the first, signaling a breakdown of our perturbation scheme. This will occur when $E\sim r_+^2 \sim T_H(0)^2$. This is an effect that can only be seen at third order in perturbation theory. 
 It is interesting and nontrivial that the full time-dependence of both the current and stress tensor can be expressed in terms of a single effective time-dependent temperature. 

So far we have seen that for $D = 3$, the current and stress tensor acquire a time dependence that can be reabsorbed in terms of an effective temperature. However, one of the most striking things that happens in $D=4$ is the fact that the system is isotropic. Obviously we cannot test if this is a special property of $D=4$ by studying a system in $D=3$! We will next briefly discuss the situation in five dimensions.

\subsection{Nonlinear $DC$ conductivity in $D = 5$}
The computation of the five-dimensional $DC$ conductivity is more involved. It follows very similar steps to those outlined in the previous section, so we will just present the final results and point out where they differ from the three-dimensional case. More details can be found in the Appendix. The first difference occurs at second order: we find one more variable, $\Phi_2$, that has a different form from $\Phi_1$, indicating that isotropy is no longer maintained. This is best seen if we look at the holographic stress energy tensor extracted at infinity. Before proceeding, let us just note that in addition to the usual counter terms presented in \cite{Balasubramanian:1999re} there is an additional one to renormalize the logarithmic divergences in the action. The full set of counterterms is given by
\begin{equation}
4\pi G\,S_c = -2\int_{\partial\mathcal{M}} \dd x^4\sqrt{-\gamma}\left(1-\frac{\tilde{R}}{4}\right)+\log\left(\frac{r_\star}{r_{\Lambda}}\right)\left.\int_{\partial\mathcal{M}} \dd x^4\sqrt{-\gamma}\tilde{F}^{\mu\nu}\tilde{F}_{\mu\nu}\right|_{r= r_{\Lambda}}
\label{eq:counter}
\end{equation}
where $\tilde{R}$ is the boundary Ricci scalar, $\tilde{F}_{\mu\nu}$ the components of the boundary Maxwell field strength and $r_\star$ just parametrizes the ambiguity in picking an energy scale for the logarithmic renormalization. The only nonzero components of the stress energy tensor turn out to be
\begin{subequations}
\begin{align}
&4\pi G\,T_{tt} = \frac{3 r_+^4}{4}+\frac{E^2}{4}\left[4 r_+ t - \pi+1+2 \log\left(\frac{r_\star}{r_+}\right)\right] \label{eq:stressenergy5Da}\,,\\
&4\pi G\,T_{xx} = \frac{r_+^4}{4}+\frac{E^2}{12}\left[4 r_+ t-\pi +3+4\log 2-6\log\left(\frac{r_\star}{r_+}\right)\right] \label{eq:stressenergy5Db}\,,\\
&4\pi G\,T_{y_1y_1}=4\pi G\,T_{y_2y_2}=\frac{r_+^4}{4}+\frac{E^2}{12} \left[4 r_+ t-\pi +3-2\log 2+6\log\left(\frac{r_\star}{r_+}\right)\right]\label{eq:stressenergy5Dc}\,.
\end{align}
\end{subequations}
The fact that $T_{xx}\neq T_{y_1y_1}$ shows that the solution is anisotropic, in contrast with $D=4$. Notice that the difference $T_{xx} -T_{y_1y_1}$ is time independent, and depends on the temperature of the initial black hole before the electric field is added, $T_{0} = r_+/\pi$. For low initial temperatures, $T_{xx} < T_{y_1y_1}$, while for high initial temperatures, the situation is reversed. Only for one particular temperature, $T_{0} = \sqrt 2 r_\star/\pi$ do the pressures remain isotropic.

In terms of the effective  temperature (\ref{eq:effectivetemperature}), 
this stress tensor can be written:
\be
4 \pi G \, T_{\mu\nu} = \frac{\pi^4}{4} T_H(t)^4 \Delta_{\mu\nu} + E^2 \tau_{\mu\nu},
\ee
where the first term is the usual expression for the CFT$_4$ dual to pure Einstein gravity in thermal equilibrium at a temperature $T_H(t)$, with $\Delta_{\mu\nu} \equiv \mbox{diag}(3,1,1,1)$, and the second term indicates a time-independent, anisotropic correction. The thermal contribution remains isotropic as expected.

We can also proceed to third order, and compute the resulting current as a function of $E^2$. As expected from the Ward identities, it is time dependent:
\begin{equation}
4\pi G\,J = E\,r_+\left[1+\frac{E^2}{12 r_+^4}\left(4 r_+ t -1-\pi+\log 2+4\log^2 2\right)\right]\,.
\end{equation}
In terms of the effective  temperature (\ref{eq:effectivetemperature}), the conductivity to order second order in $E$ reduces to:
\begin{equation}
4\pi G\,\sigma_{DC}=\pi\,T_H(t)\left[1+\frac{E^2}{6\,r_+^4}(2 \log^22-1)\right]+\mathcal{O}(E^4)\,. 
\label{eq:conductivity5D}
\end{equation}
As in the three-dimensional case, the time dependence in both the conductivity and the stress tensor can be reabsorbed into the effective temperature.

\section{Discussion}

We now attempt to interpret our results from the point of view of the field theory. Our discussion owes much to \cite{Karch:2010kt, 2005cond.mat..1758G}. Consider a general relativistic CFT in $D-1$ spacetime dimensions starting in the vacuum and apply an electric field source $E$; it appears there is only a single scale in the problem, and so from dimensional analysis alone one might expect a current of the form:
\be
J(E) \sim E^{\frac{D-2}{2}}. \label{funnypow}
\ee

\begin{figure}[ht]
\centering
\includegraphics[width=.4\textwidth]{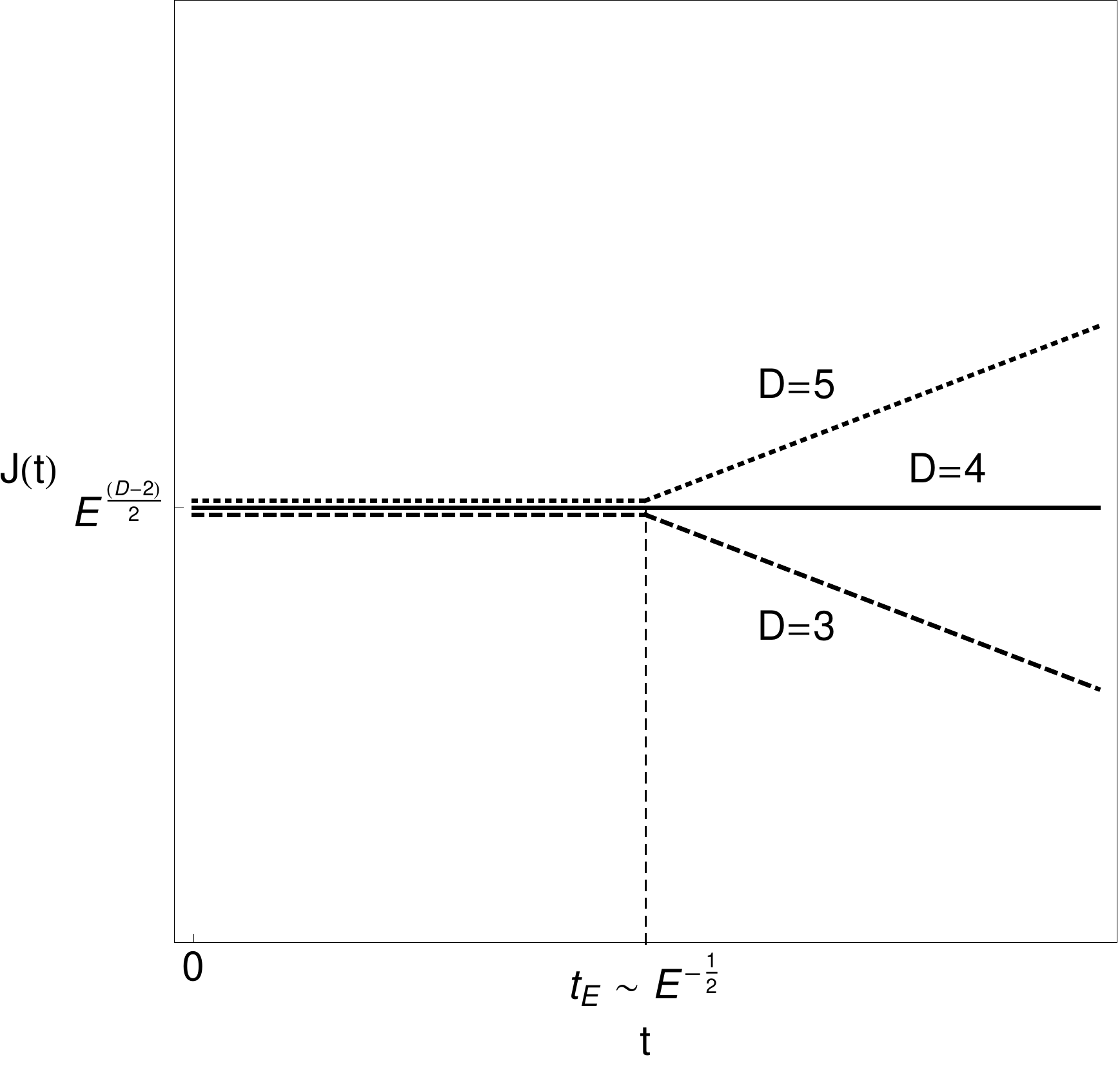}
\caption{Cartoon sketch of the expected dependence of current with time at the nonlinear level. For $t < t_E$ the current is static and determined solely by the electric field. For $t > t_E$ the effective temperature due to Joule heating is more important than the electric field. The state is now effectively thermal, and the current is determined by the effective temperature, which always grows in time but affects the system differently depending on the mass dimension of the conductivity. We find that for $D = 4$ the current is still static.}
\label{fig:nonlinearresp}
\end{figure}

However this is a little too fast. We know that the system will eventually begin to heat up via Joule heating, introducing a new scale, the effective temperature $T(t)$. This effective temperature will become comparable to the energy in the current after a time $t_E \sim E^{-\frac{1}{2}}$, after which the dominant scale in the problem is the temperature and not the electric field (see Fig. \ref{fig:nonlinearresp}). Thus while the scaling \eqref{funnypow} should apply for $t \ll t_E$, for $t \gg t_E$, we now expect a current of the form
\be
J(E) \sim \sigma(T(t),E) E, \label{dissipative}
\ee
where $\sigma(T,E)$ is a nonlinear conductivity that depends on the effective temperature. Unlike \eqref{funnypow}, this function should have a smooth limit as $E \to 0$, where it should be closely related to the usual DC conductivity $\sigma_{DC}(T)$ computed from linear response. The dynamics is controlled by the scattering of the quasiparticles off of this effective heat bath. 

Let us now understand how these facts relate to our gravity results. In $D = 3$ and $D = 5$, we perturb around a finite-$T$ state and so we are always in the regime governed by \eqref{dissipative}. In \eqref{eq:conductivity3D} and \eqref{eq:conductivity5D} we have shown how the time dependence of the current  can be understood completely in terms of a time-dependent effective temperature which changes in the way one expects from Joule heating, as indicated in \eqref{dissipative}. Note that in these dimensions our gravitational treatment cannot access the intrinsically nonlinear regime governed by \eqref{funnypow}, as our perturbation analysis breaks down as we take the starting temperature to zero. It would be very interesting to understand the bulk gravitational solution corresponding to \eqref{funnypow}, which should exist only for a finite amount of (boundary) time $t_E$, after which it should cross over to a conventional black hole similar to those we have studied.

We now discuss the interestingly different case of $D = 4$. Here our treatment is exact and we can access both regimes. Fascinatingly, we find no crossover between these two regimes at all: this is possible only because the exponent in \eqref{funnypow} corresponds to a linear dependence on $E$. Even the prefactors are equal, and the current does not seem to care that the black hole is heating up. 

We may make a more powerful statement still: if we allow both a starting background temperature $T_0$ and a frequency $\om$ for the applied electric field, the analog of \eqref{funnypow} is
\be
J(E,T_0,\om) \sim E^{\frac{D-2}{2}}Y\left(\frac{\om}{T_0},\frac{T_0}{\sqrt{E}}\right),
\ee
with $Y$ a dimensionless function of its arguments. We have found that for field theories dual to the Einstein-Maxwell sector of gravity on AdS$_4$, the function $Y$ does not depend on either of its arguments and is simply a constant $Y = 1$. The fact that the linear conductivity does not depend on frequency is well-known \cite{Herzog:2007ij}, and our calculation is an extension of this result to the nonlinear level. A similar result to ours was obtained in a different probe-brane setup with no backreaction in \cite{Karch:2010kt}.  

The simplicity of $Y$ is related to electromagnetic duality in the bulk \cite{Herzog:2007ij,Hartnoll:2007ip}. 
One first notes that before we apply an electric field, our background is isotropic, so the conductivities $\sigma_{xx}$ and $\sigma_{yy}$  must be equal. One then uses the fact (\ref{conductivity}) that  $\sigma_{xx}$ is computed holographically in terms of the asymptotic value of the ratio $F_{xr}/F_{xt}$. Under electromagnetic duality, this ratio becomes $F_{yt}/F_{yr}$, so (dropping numerical factors) $J_x$ becomes $E_y$ and $E_x$ becomes $J_y$. The net result is   $\sigma _{xx} \sim 1/ \sigma_{yy}$. Since the two conductivities must be equal, they must be constant. Since this argument only involves the asymptotic behavior of the Maxwell field, it applies at the full nonlinear level, and not just for linear perturbations.

We have not yet discussed a microscopic mechanism in the field theory that can result in a constant current. We expect the electric field to create massless particles via the Schwinger process; apparently the resulting increase in current is precisely balanced by the degradation due to their scattering. An example of this balance is worked out in a field-theoretical model (neglecting Joule heating) in \cite{2005cond.mat..1758G}. Such a mechanism is required to create a steady state such as \eqref{funnypow}, but one might have expected it to hold only for $t \ll t_E$. In our model this appears to hold for all times.  

We turn now to a different and more surprising fact: the pressures in our fully backreacted solution are isotropic $T_{xx} = T_{yy}$, even though the current flows only in the $x$ direction. As we discussed in section 2, from the bulk viewpoint, this is yet another consequence of electromagnetic duality; under $F \to * F$, the metric remains invariant yet the electric field and current rotate to point in the $y$ direction, requiring the stress-tensor to be isotropic. 

From the field theory  viewpoint this isotropy is  more perplexing. We offer the following heuristic explanation. Electric-magnetic duality in the bulk is related to interchanging particles with vortices in the boundary: to be more precise, the operation interchanges an ordinary current with a topological current, and thus can be understood as taking a theory of dynamical particles to a (generally different) theory of dynamical vortices \cite{Witten:2003ya}. In weak-coupling descriptions particles and vortices behave differently, and this operation is then not a self-duality. However if the theory {\it is} self-dual, as in the case studied here, this implies the existence of ``vortex''-like microscopic excitations whose couplings and strength of interactions are in some sense exactly equivalent to that of the particles carrying the current. As the particles flow parallel to the applied electric field, these vortices flow {\it perpendicularly} to it, generating a precisely counterbalancing pressure in the orthogonal direction. It appears difficult to make these notions any more precise, as the notion of massless ``vortex'' degrees of freedom coexisting with massless particle degrees of freedom necessitates a strongly coupled description. 

Independent of the microscopic explanation, the fact that the isotropy is predicted by our simple holographic model suggests that it might hold for a wide class of $2+1$ relativistic conformal field theories with a particle-vortex duality. We note that there is even experimental evidence for such a duality in the nonlinear response near the critical point of a quantum-hall/insulator transition \cite{1996Sci...274..589S}.

One does not actually need strict particle-vortex duality to get constant optical conductivity and isotropic pressures. We have argued that these follow from electromagnetic duality in the bulk. However it is clear that they also hold in some situations where there is no electromagnetic duality.
As pointed out in \cite{Karch:2010kt} (in the context of generalizations of the DBI action), if we add any polynomial function of $F_{\mu\nu} F^{\mu\nu}$ to our Maxwell action, $F_{xv} = E(v)$ remains a solution. The theory is no longer invariant under  $F\ra *F$, but the conductivity is still constant.
Furthermore, since our solution has $F^2 = 0$, the stress tensor is unchanged and the Vaidya metric (\ref{vaidya}) remains a solution of the fully backreacted  problem. Hence the pressures remain isotropic.

Many other modifications of the theory could be considered that violate electromagnetic duality. One that has been studied in the literature is the addition of terms of the form $\gamma RFF$ where $\gamma$ is an overall coefficient and $R$ denotes the curvature \cite{Myers:2010pk,Ritz:2008kh}.  One particular combination of such terms still yields second order equations of motion since it can be viewed as Kaluza-Klein reduction of Gauss-Bonnet gravity in five dimensions.
With these terms included it has been shown that the optical conductivity now depends on the frequency. We have  studied the effects of these terms perturbatively in $\gamma$ (but keeping $E$ finite) in our 4D solution and checked that isotropy is also violated. Curiously enough, this does not happen at first order in $\gamma$ but does happen at second order. It remains an open question to find the most general condition under which one has  constant optical conductivity and isotropic pressures. 
 
 Finally, another direction for further research is to study  Hawking radiation in our expanding black hole  background. In \cite{Sonner:2012if} (see also further analysis in \cite{Kundu:2013eba}) it was shown from a fluctuation analysis in a related probe-brane calculation that the effective non-equilibrium noise spectrum is thermal with an effective temperature depending on the applied field. It would be very interesting to understand the generalization of that result to our (continuously changing) background. 
 
 In general we hope that the simplicity of these analytic solutions can provide a starting point for further study of nonlinear and non-equilibrium transport.

\vskip 1cm
\centerline{\bf Acknowledgements}
\vskip 1cm

It is a pleasure to thank E. Fradkin, A. Green, S. Hartnoll, M. Kleban, D. Marolf, R. Myers, M. Rangamani and S. Sachdev for discussions. This work was supported in part by the National Science Foundation under Grant No. PHY12-05500. NI is supported in part by the National Science Foundation under Grant No. PHY11-25915.

\appendix
\section{Metric and Gauge Field Functions in $5D$}
For completeness we present in the appendix the metric and gauge field functions to third order in $\mathcal{E}$. The metric functions read:
\begin{subequations}
\begin{align}
&f^{(2)}(v,r) = -\frac{r_+}{3 r^3} \left\{\left(\frac{r^4}{r_+^4}+1\right) \arctan\left(\frac{r}{r_+}\right)+\frac{r}{r_+} \left[\frac{3 r^2}{r_+^2}+\log\left(\frac{r^2}{r_+^2}+1\right)-\frac{\pi }{2}+2\right]+2\right\}\nonumber\\
&\qquad\qquad\qquad\qquad\qquad + \frac{\left(r^4+r_+^4\right) (7+\log 2)}{6 r^3 r_+^3}+\frac{\left(r^4-2 r_+^3 r+r_+^4\right) \left(8 r_+ v-\pi\right)}{12 r^3 r_+^3}\\
&\Phi_1^{(2)}(v,r)=\frac{8 r_+ v-\pi +14+2 \log 2}{12 r r_+^3}+\frac{\pi  (3+2 \pi )}{18 r_+^4}+\frac{(1-\log 16)}{12 r_+^4}\log \left(1-\frac{r_+^4}{r^4}\right)+\frac{1}{3 r_+^4}I_1\left(\frac{r}{r_+}\right)\\
&\Phi_2^{(2)}(v,r)=-\frac{1} {4 r_+^4}\left\{\log \left[\frac{r_+^2+r^2}{\left(r_++r\right)^2}\right]-\frac{3 r_+}{r}+2 \left(\frac{r_+}{r}+1\right) \arctan\left(\frac{r}{r_+}\right)-\pi\right\}-\frac{\Phi_1^{(2)}(v,r)}{2}\nonumber \\
&\qquad\qquad\qquad\qquad\qquad +\frac{8 r_+ v-\pi +\log (4)}{8 r r_+^3}\,,
\end{align}
\end{subequations}
where
\begin{multline}
I_1(x)=2 \text{Li}_2\left(-x^2\right)-\text{Li}_2\left(\frac{2}{x^2+1}\right)+\log \left(x^2-1\right) \log \left(x^2+1\right)\\
+\frac{1}{4} \log \left[\frac{x^4 (x+1)^3}{(x-1)\left(x^2+1\right)^3}\right]-\frac{2}{x}-\frac{(x+1) \arctan x}{x}\,
\end{multline}
and $\text{Li}_2(x)$ is the second order polylogarithm function.

The gauge field functions read:
\begin{subequations}
\begin{align}
&Q^{(3)}_1(v,r)=\frac{r \left(r-r_+\right)}{3 r_+^3 \left(r+r_+\right) \left(r^2+r_+^2\right)}-\frac{4 \pi }{3}+\frac{3 r^2+2 r_+ r+r_+^2}{2 \left(r+r_+\right)^2 \left(r^2+r_+^2\right)^2}I_4\left(\frac{r}{r_+}\right)\,,\\
&\nonumber\\
&Q^{(3)}_2(v,r)=-\frac{r}{36 r_+^3 \left(r_+^4-r^4\right)}\left[39+2 (\pi -6) \pi +3 \log 2 (3+2 \log 2)\right]\nonumber \\
&-\frac{\pi  r^2}{24 r_+^4 \left(r_+^4-r^4\right)}+\frac{r}{r_+^3 \left(r_+^4-r^4\right)}I_2\left(\frac{r}{r_+}\right)+\frac{1}{r_+^6}I_3\left(\frac{r}{r_+}\right)\nonumber\\
&+\frac{r_+\,v}{3} \left\{\frac{3 r}{r_+^3 \left(r_+^4-r^4\right)}+\frac{r_+}{\left(r^3+r_+ r^2+r_+^2 r+r_+^3\right) \left(r_+^4-r^4\right)}\left[1-\frac{r}{r_+}\left(\frac{3r^2}{r_+^2}+\frac{3 r}{r_+}+7\right)\right]\right\}\,,
\end{align}
\end{subequations}
where
\begin{align}
&I_2(x)=-\frac{2}{3} \text{Li}_2\left(-x^2\right)-\frac{1}{24 x (x+1) \left(x^2+1\right)}\Bigg\{-8 x (x+1) \left(x^2+1\right) \text{Li}_2\left(\frac{1}{2}-\frac{x^2}{2}\right)\nonumber\\
&+2 x \Big[2x \left(x^2+x+1\right)(1+\log 256) \log x+2\log 256 \log x+x \left(x^2+x+1\right) \log \left(x^2+\frac{1}{x^2}+2\right)\nonumber\\
&+2(x+1) \left(x^2+1\right) \log \left[\frac{1}{4}\left(x^2+1\right)\right]\log \left(x^2+1\right)-2\log \left(x^2+1\right)\nonumber \\
&-4 (x+1) \left(x^2+1\right) \log (x+1)\Big]+4 \left[3 x\left(x^2+x+1\right)-1\right] \arctan x+\pi  [x (3 x^2+3x+11)-1]\nonumber \\
&-2 x [x(29 x+x \log 8+17+\log 8)+25+\log8]+6+\log 4\Bigg\}\,,
\\
&
\\
&I_3(x)=\frac{x}{24 (x-1) \left(x^3+x^2+x+1\right)^2}\Bigg\{-8 (x-3) x^2+\pi  (x+1) (x+2) \left(x^2+1\right)\nonumber \\
&+12\,x^4 \mathrm{arccoth}\,x^2-2 \log \left(x^3+x^2+x+1\right)+2 \Big[3 x^4 \log (x-1)\nonumber \\
&-2x\left(2 x^3+x^2+x+1\right) \log \left(x^2+1\right)+x(5 x^3+4x^2+4x+4) \log (x+1)\nonumber\\
&-2 (x+1)^2 \left(x^2+1\right) \mathrm{arctanh}\,x\Big]+8 (x+1)+6 \log (x-1)+12\mathrm{arccoth}\,x\Bigg\}\,,
\\
\text{and}
\\
&I_4(x)=\log \left(\frac{x-1}{\sqrt{x^2+1}}\right)-\frac{\left(x^3+x^2+x+1\right)^2}{6 [x (3x+2)+1]} \log \left[\frac{(x-1)^3 (x+1)^5}{\left(x^2+1\right)^4}\right]+\mathrm{arccoth}\,x\nonumber\\
&+\frac{\left(x^3+x^2+x+1\right)^2 \arctan x}{9 x^2+6 x+3}-\frac{[x (x^2+2x+3)+4] x^3 \mathrm{arccoth}\,\left(x^2\right)}{x (3 x+2)+1}\,.
\end{align}

\bibliographystyle{JHEP}
\bibliography{nonlinear}

\providecommand{\href}[2]{#2}\begingroup\raggedright\begin{thebibliography}{10}

\bibitem{Herzog:2007ij}
C.~P. Herzog, P.~Kovtun, S.~Sachdev, and D.~T. Son, {\it {Quantum critical
  transport, duality, and M-theory}},  {\em Phys.Rev.} {\bf D75} (2007) 085020,
  [\href{http://xxx.lanl.gov/abs/hep-th/0701036}{{\tt hep-th/0701036}}].

\bibitem{Hartnoll:2007ip}
S.~A. Hartnoll and C.~P. Herzog, {\it {Ohm's Law at strong coupling: S duality
  and the cyclotron resonance}},  {\em Phys.Rev.} {\bf D76} (2007) 106012,
  [\href{http://xxx.lanl.gov/abs/0706.3228}{{\tt arXiv:0706.3228}}].

\bibitem{Witten:1998zw}
E.~Witten, {\it {Anti-de Sitter space, thermal phase transition, and
  confinement in gauge theories}},  {\em Adv.Theor.Math.Phys.} {\bf 2} (1998)
  505--532, [\href{http://xxx.lanl.gov/abs/hep-th/9803131}{{\tt
  hep-th/9803131}}].

\bibitem{Horowitz:1998ha}
G.~T. Horowitz and R.~C. Myers, {\it {The AdS / CFT correspondence and a new
  positive energy conjecture for general relativity}},  {\em Phys.Rev.} {\bf
  D59} (1998) 026005, [\href{http://xxx.lanl.gov/abs/hep-th/9808079}{{\tt
  hep-th/9808079}}].

\bibitem{Donos:2011bh}
A.~Donos and J.~P. Gauntlett, {\it {Holographic striped phases}},  {\em JHEP}
  {\bf 1108} (2011) 140, [\href{http://xxx.lanl.gov/abs/1106.2004}{{\tt
  arXiv:1106.2004}}].

\bibitem{Hawking:1982dh}
S.~Hawking and D.~N. Page, {\it {Thermodynamics of Black Holes in anti-De
  Sitter Space}},  {\em Commun.Math.Phys.} {\bf 87} (1983) 577.

\bibitem{Faulkner:2012gt}
T.~Faulkner and N.~Iqbal, {\it {Friedel oscillations and horizon charge in 1D
  holographic liquids}},  {\em JHEP} {\bf 1307} (2013) 060,
  [\href{http://xxx.lanl.gov/abs/1207.4208}{{\tt arXiv:1207.4208}}].

\bibitem{Balasubramanian:1999re}
V.~Balasubramanian and P.~Kraus, {\it {A Stress tensor for Anti-de Sitter
  gravity}},  {\em Commun.Math.Phys.} {\bf 208} (1999) 413--428,
  [\href{http://xxx.lanl.gov/abs/hep-th/9902121}{{\tt hep-th/9902121}}].

\bibitem{Karch:2010kt}
A.~Karch and S.~Sondhi, {\it {Non-linear, Finite Frequency Quantum Critical
  Transport from AdS/CFT}},  {\em JHEP} {\bf 1101} (2011) 149,
  [\href{http://xxx.lanl.gov/abs/1008.4134}{{\tt arXiv:1008.4134}}].

\bibitem{2005cond.mat..1758G}
A.~G. Green and S.~L. Sondhi, {\it {Nonlinear Quantum Critical Transport and
  the Schwinger Mechanism for a Superfluid-Mott-Insulator Transition of
  Bosons}},  {\em Phys. Rev. Lett.} {\bf 95} (Dec, 2005) 267001,
  [\href{http://xxx.lanl.gov/abs/cond-mat/0501758}{{\tt cond-mat/0501758}}].

\bibitem{Witten:2003ya}
E.~Witten, {\it {SL(2,Z) action on three-dimensional conformal field theories
  with Abelian symmetry}},  \href{http://xxx.lanl.gov/abs/hep-th/0307041}{{\tt
  hep-th/0307041}}.

\bibitem{1996Sci...274..589S}
D.~{Shahar}, D.~C. {Tsui}, M.~{Shayegan}, E.~{Shimshoni}, and S.~L. {Sondhi},
  {\it {Evidence for Charge-Flux Duality near the Quantum Hall
  Liquid-to-Insulator Transition}},  {\em Science} {\bf 274} (Oct., 1996)
  589--592, [\href{http://xxx.lanl.gov/abs/cond-mat/9510113}{{\tt
  cond-mat/9510113}}].

\bibitem{Myers:2010pk}
R.~C. Myers, S.~Sachdev, and A.~Singh, {\it {Holographic Quantum Critical
  Transport without Self-Duality}},  {\em Phys.Rev.} {\bf D83} (2011) 066017,
  [\href{http://xxx.lanl.gov/abs/1010.0443}{{\tt arXiv:1010.0443}}].

\bibitem{Ritz:2008kh}
A.~Ritz and J.~Ward, {\it {Weyl corrections to holographic conductivity}},
  {\em Phys.Rev.} {\bf D79} (2009) 066003,
  [\href{http://xxx.lanl.gov/abs/0811.4195}{{\tt arXiv:0811.4195}}].

\bibitem{Sonner:2012if}
J.~Sonner and A.~G. Green, {\it {Hawking Radiation and Non-equilibrium Quantum
  Critical Current Noise}},  {\em Phys.Rev.Lett.} {\bf 109} (2012) 091601,
  [\href{http://xxx.lanl.gov/abs/1203.4908}{{\tt arXiv:1203.4908}}].

\bibitem{Kundu:2013eba}
A.~Kundu and S.~Kundu, {\it {Steady-state Physics, Effective Temperature
  Dynamics in Holography}},  \href{http://xxx.lanl.gov/abs/1307.6607}{{\tt
  arXiv:1307.6607}}.

\end{thebibliography}\endgroup

\end{document}